\PassOptionsToPackage{svgnames}{xcolor}
\documentclass[submission,copyright,creativecommons]{eptcs}
\providecommand{\event}{VPT 2019} % Name of the event you are submitting to
\usepackage{underscore}           % Only needed if you use pdflatex.
\usepackage{listings}
\usepackage{tikz}
\makeatletter
\newcommand*\idstyle{%
        \expandafter\id@style\the\lst@token\relax
}
\def\id@style#1#2\relax{%
        \ifcat#1\relax\else
                \ifnum`#1=\uccode`#1%
                        \ttfamilywithbold\bfseries
                \fi
        \fi
}
\makeatother

\definecolor{darkRed}{RGB}{100,0,10}
\definecolor{darkBlue}{RGB}{10,0,100}
\newcommand*{\ttfamilywithbold}{\ttfamily}

\lstdefinelanguage{FortyTwo}[]{Java}{morekeywords={%
  M,
  exception,error,mut,imm,
  read,capsule,lent,assert
  with,in,immutable,trait,using,
  on,var,loop,reuse,method,is
  },
   basicstyle=\ttfamily,
   keywordstyle=\ttfamilywithbold\bfseries\color{darkRed},
   identifierstyle=\idstyle,
   showstringspaces=false,
   mathescape=true,
   xleftmargin=0pt,
   xrightmargin=0pt,
   breaklines=false,
   breakatwhitespace=false,
   breakautoindent=false,
   tabsize=2,
   commentstyle=\color{darkBlue}\ttfamily,
   stringstyle=\color{darkRed}\ttfamily,
   literate=
                 {\%}{{\mbox{\textbf{\%}}}}1
                 {~} {$\sim$}1
 }

\usepackage{xcolor}

\usepackage{verbatim}

\title{Iteratively Composing Statically Verified Traits}
\def\titlerunning{Iteratively Composing Statically Verified Traits}

\RequirePackage{array}
\newenvironment{authors}[1]%
  {\begingroup
   \gdef\estyle{}%
   \renewcommand\institute[1]%
     {\\\multicolumn{#1}{@{}c@{}}{\scriptsize\begin{tabular}[t]{@{}>{\footnotesize}c@{}}##1\end{tabular}}}%
   \renewcommand\email[1]%
     {\gdef\estyle{\footnotesize\ttfamily}\\##1\gdef\estyle{}}
   \begin{tabular}[t]{@{}*{#1}{>{\estyle}c}@{}}
  }%
  {\end{tabular}%
   \endgroup
  }

\def\vuw{\institute{School of Engineering and Computer Science\\%
			Victoria University of Wellington\\%
			Wellington, New Zealand}}
\author{
	\begin{authors}{4}
	Isaac Oscar Gariano & Marco Servetto & Alex Potanin & Hrshikesh Arora
	\vuw
	\email{~&\hspace{-6pt}\{isaac, marco.servetto,&  \hspace{-13pt} alex, arorahrsh\}@myvuw.ac.nz &~}
	\end{authors}
}

\def\authorrunning{I.\,O. Gariano, M. ,Servetto, A. Potanin \& H. Arora}

\chardef\Slash=`\/
\catcode\Slash=\active
\chardef\other=12 % char code for other characters
\def/#1/{%
	\ifx/#1/% #1 is empty
		\Slash% just print a slash
	\else%
		\lstinline/#1/%
	\fi%
}
\let\oldinput=\input
\def\input#1{\oldinput{\detokenize{#1}}} % Don't expand commands in input (needed since / is a command)

\makeatletter
\renewcommand*\idstyle{%
	\expandafter\id@style\the\lst@token\relax}
\def\id@style#1#2\relax{%
        \ifcat#1\relax\else
                \ifnum`#1=\uccode`#1%
                        \color{DarkGreen}%
                \fi%
        \fi%
}
\makeatother

\begin{document}
\maketitle
\begin{abstract}
Static verification relying on an automated theorem prover can be very slow and brittle: since static verification is undecidable, correct code may not pass a particular static verifier.
In this work we use metaprogramming to generate code that is correct by construction.
A theorem prover is used only to verify initial ``traits'': units of code that can be used to compose bigger programs.

In our work, meta-programming is done by trait composition, which starting from correct code, is guaranteed to produce correct code.
We do this by extending conventional traits 
with pre- and post-conditions for the methods; we also extend the  traditional trait composition (/+/) operator to check the compatibility of contracts. In this way, there is no need to re-verify the produced code.

We show how our approach can be applied to the standard ``power'' function example, where metaprogramming generates optimised, and correct, versions when the exponent is known in advance.
\end{abstract}

\section{Introduction}
With this short paper we contribute
to research on safe metaprogramming, showing how combining 
pre//post conditions,
trait composition and metaprogramming
it is possible to create metaprograms that generate code which is correct by construction.
That is: only the original source code itself needs to be verified (for example by a theorem prover), and not the code produced by metaprogramming.
This is important since the original code is often orders of magnitude smaller than the generated code.
We start by providing some background on those three research areas: pre//post conditions, traits, and metaprogramming.

\textbf{Pre//post conditions}:
object-oriented (OO) languages supporting static verification usually extend the syntax for method declarations
to support \emph{contracts} in the form of pre- and post-conditions~\cite{Meyer:1988:OSC:534929}.
Correctness is defined only for code annotated with such contracts.

We say that a method is \emph{correct}, if whenever its precondition holds on entry, the precondition of every directly invoked method holds, and the postcondition of the method holds when the method returns. Automated static verification typically works by asking an automated theorem prover to verify that each method is correct individually, by assuming the correctness of every other method~\cite{barnett2004spec}. This process can be very slow and can produce unexpected results: since static verification is undecidable, correct code may not pass a particular static verifier.
Many static verification approaches are not resilient to
standard refactoring techniques like 
method inlining. Sometimes static verification even times out, making the behaviour even more sensitive to such refactoring techniques.

\textbf{Traits}: originally introduced in Smalltalk by Scharli~\cite{scharli2003traits}, traits are 
units of code reuse. They were created as a simpler way of performing multiple inheritance without the usual complexity.
Traits are just a set of method declarations.
Such methods can be abstract and
mutually recursive by using the implicit /this/ parameter.
Traits are different from Java-style abstract classes as
%\begin{itemize}
%\item
they are only for reuse: trait names do \emph{not} define types.
%\item Traits contains only methods; thus they have no field or constructors.
%\item
Trait composition is seen as a form of flattening: after the composition, the resulting code contains copies of the adapted methods from the original traits, but do not reference the original traits directly. Since there is no trace of
the point of origin of the code, /super/ calls are not 
directly available and need to be somehow emulated.
%\item
Traits can be combined using many different composition operators, not just \emph{extends}.
%\end{itemize}
In this work we will rely on the traditional composition operators /+/(plus), /rename/ and /hide/.

The /+/ operator is the main way to compose traits
~\cite{scharli2003traits,LagorioSZ09}.
The result of /+/ will contain all the methods from both operands. 
Crucially, it is possible to sum traits where a method is declared in both operands; in this case at least one of the two competing methods needs to be abstract, and the signatures of the two competing methods need to be \emph{compatible}.
In this way, /+/ has an expressive power similar to multiple inheritance.

\emph{Rename} and \emph{hide} adapt a single trait by renaming a method or by making a method private.
Many works in the literature allow adapting traits by renaming or hiding methods~\cite{servetto2014meta,reppy2007metaprogramming,liquori2008feathertrait}. Hiding a method may also trigger inlining if the method body is simple enough or used only once.

Consider the following example code, where we use those 3 operators:
\begin{lstlisting}
Trait a=class{Int hello(){return 1;}}
Trait b=class{
  abstract Int hello();
  String world(){return "["+this.hello()+"]";}}
Trait c=(a+b)[hide hello()][rename world()->hello()]
\end{lstlisting}
After flattening we would get the following result; where /c/ now contains a single method called
/world()/, with the body of the method originally called /hello()/ declared in trait /b/. This body contains the
inlined version of method /a.world()/ that has been hidden.
Note how the order of operations is important.
\begin{lstlisting}
Trait a=/*as before*/
Trait b=/*as before*/
Trait c=class {String world(){return "["+1+"]";}}
\end{lstlisting}

\textbf{Metaprogramming} is often used to programmatically generate faster specialised code when some parameters are known in advance, this is particularly useful where the specialisation mechanism is too complicated for a generic compiler to automatically derive~\cite{Ofenbeck:2017:SGP:3136040.3136060}.
A \emph{metaprogram} is a program, method or function 
that produces code. Depending on the kind of metaprogramming, such code can be directly executable or can be just an abstract representation of behaviour.
Metaprogramming is called \emph{metacircular} when the language used to write the metaprogram is the same language of the generated code.
Metaprogramming can happen at run time, or at compile time. In the latter case, the produced code can be needed to typecheck and compile the rest of the code in the program.
In this paper we will rely on 
Iterative Composition: 
a metacircular metaprogramming technique relying on \emph{compile-time execution} (a form of execution also used by~\cite{sheard2002template}).
This disciplined form of metaprogramming introduced by Servetto and Zucca \cite{servetto2014meta}, is based on the trait composition operators described before, but lifted at the expression level.
This means that arbitrary expressions can be used as the right hand side of trait and class declarations; during compilation such expressions will be evaluated to produce a /Trait/, which provides the body of the class. In this way metaprograms can be represented as otherwise normal functions//methods that return a /Trait/, without requiring the use of any additional `metalanguage'.

\section{Combining Metaprogramming and Static Verification}

A na\"ive way to combine metaprogramming and static verification could be to use metaprogramming to generate code together with contracts, and then once the metaprogramming has been run,
% \MSDel{ensure the correctness of} 
statically verify the resulting code. 
However, the resulting code could be much larger than the input to the metaprogramming, and so it could take a long time to statically verify.
Moreover, one of the many goals of metaprogramming is to make it easier to generate many specialised versions of the same  code.
%, Even if the generated code was produced by using straightforward transformations and compositions over the input code, a SV might not verify it's correctness.
The aim of our work is to statically verify only the original source code itself, and not the code produced by metaprogramming.
Instead, we
ensure that the result of metaprogramming is correct by construction.

%Here we use the disciplined form of metaprogramming introduced by Servetto and Zucca \cite{servetto2014meta}, which is based on trait %composition and adaptation~\cite{scharli2003traits}.
%Here a /Trait/ is a unit of code: a set of method declarations.
%Such methods can be abstract and can be
%mutually recursive by using the implicit parameter /this/.

We extend~\cite{servetto2014meta} by allowing
methods to be annotated with pre//post-conditions.
In addition to requiring that all the traits are well-typed before they are used (as in~\cite{servetto2014meta}) we also require that traits are correct in terms of their method contracts.
/Trait/s directly written in the source code are statically verified, while traits resulting from metaprogramming are ensured correct by only providing trait operations that preserve correctness. 
In particular, we \textbf{only} need to extend the checking performed by the traditional trait composition (/+/)  operator to also check the compatibility of contracts.

%
%The result of composing and adapting /Trait/s is also correct and well-typed.
%
Our metaprogramming approach does not allow generating code from scratch, such as by directly generating ASTs, rather the language provides a specific set of primitive composition and adaptation operators which preserve correctness.
Thus the result of metaprogramming is guaranteed to be well typed and correct.
%Note that generated code may not be able to pass a particular static verifier.

Static verification usually handles /extends/ and /implements/ by verifying that every 
time a method is implemented//overriden, 
the Liskov substitution principle~\cite{Liskov:1994:BNS:197320.197383} is satisfied
by checking that the contracts of the method in the derived class implies the contract of any corresponding methods in its base classes. 
 In this way, there is no need to re-verify
inherited code in the context of the derived class.
This concept is easily adapted
to handle trait composition, which simply provides another way to implement an /abstract/ method.
When traits are composed,
it is sufficient
to match the contracts of the few composed methods
to ensure the whole result is correct.

\section{Concrete Example}

In our example below we will use the notation /@requires($predicate$)/ 
for specifying a precondition and /@ensures($predicate$)/ 
to specify a postcondition; where $predicate$ is a boolean expression
in terms of the parameters of the method (including /this/), and for the /@ensures/ case, the /result/ of the method.
Suppose we want to implement an efficient exponentiation function, we could use recursion and the common technique of `repeated squaring':
\vspace{-1ex}
\begin{lstlisting}
@requires(exp > 0)
@ensures(result == x**exp)//Here x**y means x to the power of y
Int pow(Int x, Int exp) {
	if (exp == 1) return x;
	if (exp %2 == 0) return pow(x*x, exp/2); // exp is even
	return x*pow(x, exp-1); }  // exp is odd
\end{lstlisting}
If the exponent is known at compile time,
unfolding the recursion produces even more efficient code:
\vspace{-1ex}
\begin{lstlisting}
@ensures(result == x**7) Int pow7(Int x) { 
  Int x2 = x*x; // x**2
  Int x4 = x2*x2; // x**4
  return x*x2*x4; } // Since 7 = 1 + 2 + 4
\end{lstlisting}
\vspace{-1ex}

We now show how \emph{Iterative Composition} %(introduced in~\cite{servetto2014meta} and
(enriched by the contract compatibility check we proposed) % performing in trait composition) 
can be used to write a metaprogram that given an exponent, produces code like the above.
%Iterative Composition is a metacircular metaprogramming technique relying on \emph{compile-time execution} (a form of execution also used by~\cite{sheard2002template}),
%in our context this means that arbitrary expressions can be used as the right hand side of a class declaration; during compilation such expressions will be evaluated to produce a /Trait/, which provides the body of the class. In this way metaprograms can be represented as otherwise normal functions//methods that return a /Trait/, without requiring the use of any additional `metalanguage'.
 
%\vspace{-1ex}
First we will define tree traits: /base/, /even/ and /odd/.
\begin{lstlisting}
Trait base=class {//induction base case: pow(x)==x**1
  @ensures(result>0) Int exp(){return 1;}  
  @ensures(result==x**exp()) Int pow(Int x){return x;}
}
Trait even=class{//if _pow(x)==x**_exp(), pow(x)==x**(2*_exp())
  @ensures(result>0) Int $\_$exp();
  @ensures(result==2*$\_$exp()) Int exp(){return 2*$\_$exp();}
  @ensures(result==x**$\_$exp()) Int $\_$pow(Int x);
  @ensures(result==x**exp()) Int pow(Int x){return $\_$pow(x*x);}
}
Trait odd=class {//if _pow(x)==x**_exp(), pow(x)==x**(1+_exp())
  @ensures(result>0) Int $\_$exp();
  @ensures(result==1+$\_$exp()) Int exp(){return 1+$\_$exp();}
  @ensures(result==x**$\_$exp()) Int $\_$pow(Int x);
  @ensures(result==x**exp()) Int pow(Int x){return x*$\_$pow(x);}
}
\end{lstlisting}

They are the basic building blocks we will use to compute our result. They will be compiled, typechecked and statically verified before being used in any way.
Note that we could use /base/ directly:
we could write /class Pow1: base/; this would generate a class such that /new Pow1().pow(x)==x**1/.
The other two traits have abstract methods; implementations for /$\_$pow(x)/ and /$\_$exp()/ must be provided. However, given the contract of /pow(x)/,
and the fact that /even/ and /odd/ have both been statically verified,
if we supply method bodies respecting these contracts, we will get \emph{correct} code, without the need for further static verification.
Since all occurrences of names are consistently renamed, \textbf{renaming and hiding preserve code correctness}.
Method names starting with $\_$, like /$\_$pow(x)/ and /$\_$exp()/, are not special and are not treated in any special way by trait composition. We use the $\_$ naming convention to emulate /super/ when using trait composition: a call to /$\_$exp()/ is used like a /super.exp()/ call in a language with conventional class based inheritance like Java.

The /compose(current,next)/ method starts by renaming the /exp()/ and /pow(x)/ methods of\\* /current/
so that they satisfy the contracts in /next/ (which will be 
/even/ or /odd/).
\begin{lstlisting}
//`compose' performs a step of iterative composition
Trait compose(Trait current, Trait next){
  current = current[rename exp()->$\_$exp(), pow(x)->$\_$pow(x)];
  return (current+next)[hide $\_$exp(), $\_$pow(x)];}
@requires(exp>0)//the entry point for our metaprogramming
Trait generate(Int exp) {
  if (exp==1) return base;
  if (exp%2==0) return compose(generate(exp/2),even);
  return compose(generate(exp-1),odd);
};
\end{lstlisting}
%\vspace{-1ex}

%The /+/ operator is the main way to compose traits%
%~\cite{scharli2003traits,LagorioSZ09}.
%The result of /+/ will contain all the methods from both operands. 

%Crucially, it is possible to sum traits where a method is declared in both operands; in this case at least one of the two competing methods needs to be abstract, and the signatures of the two competing methods need to be \emph{compatible}.
Then, the operator /+/ is used to compose the code of the parameters.
Here we show how we ensure that the traditional /+/ operator also handles contracts: we require that the contract annotations of the two competing methods are \emph{compatible}.
In this paper, we just require them to be syntactically identical. Relaxing this constraint is an important future work.
Thanks to this constraint \textbf{the sum operator also preserves code correctness}. %\IO{There are many variations of the /+/ operator, in particular, we could easily extend our contract matching to work with an nary operator}.

The sum is executed when the method /compose/
%\IO{\footnote{\IO{a generic implementation of this method that renames and hides conflicting methods has been implemented L42~\cite{l42}}}}
runs: if the matched contracts are not identical an exception will be raised. A leaked exception during compile-time metaprogramming would become a compile-time error. 
Our approach is very similar to~\cite{servetto2014meta} and does not guarantee the success of the code generation process, rather it guarantees that if it succeeds, correct code is generated.

Executing /compose(base,even)/ or /compose(base,odd)/ will pass this test: since the contract of /base.pow()/
is the same of /even.$\_$pow()/ and /even.$\_$pow()/, and the same for /exp()/.

Finally the /$\_$pow(x)/ and /$\_$exp()/ method are hidden, so that the structural shape of the result is
the same as /base/'s.
Note that this structural equality includes the contracts of methods.

Note that /Trait/s are first class values and can be manipulated with a set of primitive operators that preserve code correctness and well-typedness.
In this way, by inductive reasoning, we can start from the /base/ case and then recursively compose /even/ and /odd/ until we get the desired code.
Note how the code of /generate(exp)/ follows the same scheme of the code of /pow(x,exp)/ in line 1.

To understand our example better, imagine executing the code of /generate(7)/ while keeping /compose/ in symbolic form. We would get the following (where /c/ is short for /compose/):
\vspace{-1ex}
\begin{lstlisting}[numbers=none]
generate(7) == c(generate(6),odd) == ...
 == c(c(c(c(base,even),odd),even),odd)
\end{lstlisting}
\vspace{-1ex}
As /base/ represents /pow1(x)/; /c(base,even)/ represents /pow2(x)/. Then \Q@c(/*pow2(x)*/,odd)@ represents \Q@pow3(x)@, \Q@c(/*pow3(x)*/,even)@ represents \Q@pow6(x)@, and finally,
\Q@c(/*pow6(x)*/,odd)@ represents \Q@pow7(x)@.
The code of each /$\_$pow(x)/ method is only executed once for each top-level /pow(x)/ call, so the /hide/ operator can inline them.
Thus, the result could be identical to the manually optimized code in line 7.
We can use our /generate(7)/ as follows:
\begin{lstlisting}
class Pow7: generate(7)//generate is executed at compile time
//the body of class Pow7 is the result of generate(7)
/*example usage:*/
new Pow7().pow(3)==2187//Compute 3**7
\end{lstlisting}

%\IO{We are investigation how an additional check can be performed to ensure the resulting code has specific contracts. However, our approach does guarantee that the result will be correct according to whatever contracts it contains.} 
\section{Future Work}
Our approach, as presented in this short paper, only guarantees that code resulting from metaprogramming follows its own contracts, it does
not statically ensure what those contracts may be. As future work, we are investigating how the resulting contracts can be ensured to have a particular meaning or form.
To do so, we need to allow assertions on the contracts of /Trait/s to be used within pre//post conditions.
For example we could allow post conditions like\\*
%\begin{lstlisting}[numbers=none]
/@ensures(result.$\mathit{methName}$.ensures ==\ $\mathit{predicate}$)/ \\*
%\end{lstlisting}
to mean that the resulting /Trait/ has
a method
called $\mathit{methName}$, whose /@ensures/ clause is syntactically identical to  /$predicate$/; whilst
\\*
/@ensures(result.$\mathit{methName}$.ensures ==>\ $\mathit{predicate}$)/
\\*
would use a static verifier to ensure that $\mathit{methName}$'s /@ensures/ clause logically implies $\mathit{predicate}$.
With these two features we could annotate the method /generate(exp)/ in line 32 above as:
\begin{lstlisting}
@requires(exp>0)
@ensures(result.exp().ensures ==> (result==exp))
@ensures(result.pow(x).ensures == (result==x**exp()))
Trait generate(Int exp) {...}
\end{lstlisting}

\vspace{-1ex}
In this way, we could statically verify the /generate(exp)/ method, however we fear such verification will be too complex or impractical. 
We could instead automatically check the above postconditions after each call to /generate(exp)/. If /generate(exp)/ is used to define a class (such as /Pow7/ above), we will guarantee that such class has the expected contracts, before it is used. Thus
there is no need to ensure the correctness of the metaprogram itself: such runtime checks are sufficient to ensure that after compilation, the code produced by metaprogramming has its expected behaviour.
%\IODel{In this case we could defer those difficult//novel predicates to run-time checks, without losing much safety:
%Iterative Composition execute metaprogramming code at
%compile time, thus even run-time verification of metaprograms would happen at compile time. This consideration could result in a crucial design decision: code performing metaprogramming does not need to be verified by SV to produce code annotated with the desired contracts; it may be sufficient to apply some type of runtime verification during compile-time execution.} \IOComm{I did a major rewording since we actually have multiple compile-times and run-times, so your version is confusing, hopefully my version makes the point more clear.}
%For example, the following code:
%\vspace{-1ex}
%\begin{lstlisting}[numbers=none]
%@ensures(new Pow7().exp()==7&&Pow7.pow.ensures=="result==x**exp()")
%class Pow7: generate(7)
%\end{lstlisting}
%\vspace{-1ex}
%may require the static verifier to check that the execution of
%/new Pow7().exp()/ will deterministically reduce to /7/, and that the /ensures/ clause of 
%/Pow7.pow/ is syntactically equivalent to 
%/result==x**exp()/. Note how this final step of static verification does not need to re-verify the body of
%/Pow7.pow/ and only needs to do a coarse grained 
%determinism check on the implementation of /Pow7.exp()/, before symbolically executing it.

\section{Conclusion}
By exploiting conventional OO static verification techniques, we have extended the Iterative Composition form of metaprogramming with a simple contract compatibility check, to statically ensure the correctness of code produced by such metaprogramming. In particular, our approach does not require static verification of the result of metaprogramming, but only requires verification of code present directly in source code.
Following general terminology in software verficiation, we say that a trait is \emph{correct} when its methods respect their contracts.
Thus our result is that starting from a set
of well typed and correct traits, 
any code resulting from arbitrary many steps of trait composition will also be correct and well typed.
In this way, the programmer need only provide correct bulding blocks using traits;
code generated by metaprogramming can be integrated with a correct program without needing to use expensive theorem provers or manual verification.
Our example is applied to code specialization of a mathematical function, but our experience suggests that Iterative composition can be used to synthesize arbitrary behaviour.

\catcode\Slash=12% turn of my slash
\bibliographystyle{eptcs}
\bibliography{paper}
%\clearpage
%\appendix
%\input{appendix}
\end{document}